\documentclass[11pt]{article} 
\usepackage[T1]{fontenc}
\usepackage[utf8]{inputenc} 

\usepackage{geometry} 
\usepackage{graphicx} 

\usepackage{booktabs} 
\usepackage{array} 
\usepackage{paralist} 
\usepackage{verbatim} 
\usepackage{subfig} 
\usepackage{amsmath} 
\usepackage{amssymb} 
\usepackage{url} 
\usepackage{epstopdf} 
\usepackage{color}
\usepackage{stackengine}
\usepackage{graphicx}
\usepackage{cite}
\usepackage{hyperref}

\usepackage{fancyhdr} 
\usepackage{sectsty}
\allsectionsfont{\sffamily\mdseries\upshape} 

\usepackage[nottoc,notlof,notlot]{tocbibind} 
\usepackage[titles,subfigure]{tocloft} 

\title{Probing Limits of Information Spread with Sequential Seeding}
\author{Jarosław Jankowski$^{1}$\footnote{E-mail:jjankowski@wi.zut.edu.pl}, Boleslaw K. Szymanski$^{2,4}$, Przemys{\l}aw Kazienko$^{3}$, \\ Rados{\l}aw Michalski$^{3}$, Piotr Bródka$^{3}$}

\date{}
\begin{document}
\maketitle

\begin{flushleft}
$^{\bf{1}}$ Faculty of Computer Science and Information Technology, West Pomeranian University of Technology, 70-310 Szczecin, Poland\\
$^{\bf{2}}$ Social and Cognitive Networks Academic Research Center and Department of Computer Science, Rensselaer Polytechnic Institute, Troy NY 12180, USA\\
$^{\bf{3}}$ Faculty of Computer Science and Management, Wroclaw University of Science and Technology, 50-370 Wroclaw, Poland \\
$^{\bf{4}}$ Spo\l{}eczna Akademia Nauk, \L{}\'{o}d\'{z}, Poland\\
\end{flushleft}

\section*{Abstract} 
We consider here information spread which propagates with certain probability from nodes just activated to their not yet activated neighbors. Diffusion cascades can be triggered by activation of even a small set of nodes. Such activation is commonly performed in a single stage. A novel approach based on sequential seeding is analyzed here resulting in three fundamental contributions. First, we propose a coordinated execution of randomized choices to enable precise comparison of different algorithms in general. We apply it here when the newly activated nodes at each stage of spreading attempt to activate their neighbors. Then, we present a formal proof that sequential seeding delivers at least as large coverage as the single stage seeding does. Moreover, we also show that, under modest assumptions, sequential seeding achieves coverage provably better than the single stage based approach using the same number of seeds and node ranking. Finally, we present experimental results showing how single stage and sequential approaches on directed and undirected graphs compare to the well-known greedy approach to provide the objective measure of the sequential seeding benefits. Surprisingly, applying sequential seeding to a simple degree-based selection leads to higher coverage than achieved by the computationally expensive greedy approach currently considered to be the best heuristic.
\\
\newline
\textbf{Keywords:} diffusion, social network, seed selection, influence maximization, sequential seeding
\\

\section*{Introduction}
Making decisions is often difficult; this is why it is often worth to split it into consecutive stages to reduce potential risk. Such approach gives the decision maker an opportunity to learn the outcomes of the previous stages and adjust accordingly the current stage. This approach applies also to influence maximization and information spread in complex networks. One of the main challenges there is the selection for initial and activation network nodes referred to as seeds, to maximize the spread of information within the network~\cite{Kempe:2003}. 

Various factors affecting the diffusion and social influence in complex networks were analyzed including the role of different centrality measures in selection of initial influencers~\cite{Kiss2008}, impact of homophily for successful seeding~\cite{Nejad:2015} and others~\cite{liu2012seeding}. While most of the relevant research is related to marketing, the problem is more generally defined as a target set selection in combinatorial optimization in theoretical computer science~\cite{Ackerman:2010,Ben-Zwi:2011,Chiang:2013}. The influence maximization problem is also explored in physics from the perspective of network structures~\cite{Galstyan2009}. Some other studies discuss the role of communities~\cite{He2015} or propose to use optimal percolation \cite{Morone:2015}. Initial research has been carried out to identify seeds for temporal~\cite{Michalski:2014,jankowski2013} and multi-layered social networks~\cite{Michalski:2013}. Several comparative studies on seeding strategies were presented in~\cite{Hinz:2011,Libai:2013}. Since the seed selection process is NP-hard~\cite{Kempe:2003}, several heuristics have been proposed. Most of them use network structural properties like degree or eigenvector measure to rank seed candidates~\cite{Hinz:2011}.

Seeding strategies have been applied to word-of-mouth marketing, social and political campaigns and diffusion of information in social media. Vast majority of them are based on the assumption that all seeds are activated at the beginning of the process or campaign and then diffusion starts and continues naturally without any additional support \cite{Hinz:2011}. Some recent research proposes to apply adaptive approaches with two-stage stochastic model exploring the potential of neighboring nodes \cite{Seeman2013}, further extended towards more scalable approach~\cite{Horel2015}. Using some seeds after the first stage was preliminary proposed in~\cite{sela2015improving,zhang2015dynamic} and potential of multi-period spraying for routing in delay-tolerant networks was discussed as an effective solution for propagation in computer networks in~\cite{Szymanski:2010}.

On top of that, the concept of sequential seeding was introduced~\cite{jankowski2017balancing}. It takes the advantage of delayed seeding by interweaving activating seeds with diffusion. Here we firm the foundations of sequential seeding and address a number of research questions that have not been studied before~\cite{jankowski2017balancing, jankowski2017dynamic, jankowski2017seeds}. First, we prove formally and confirm experimentally on real networks that the sequential seeding is at least as good as the single stage method and under modest conditions, it is provably better. Then, we evaluate the realistic benefits of this method by comparing its capabilities to both greedy and optimal (maximal) solutions to obtain some measure of improvements provided by this approach. Finding optimal seed sets and computing the maximal coverage for various scenarios allow us to represent gains from sequential seeding as a fraction of the space between coverage of other algorithms and the maximal coverage. This has been investigated empirically for directed and undirected networks using independent cascades model~\cite{Kempe:2003}. 

\begin {table*}[!htb]
\begin{center}
\begin{tabular}{ c p{11cm} c c}
\hline

\textbf{Strategy} & \textbf{Description} & \textbf{No. of seeding stages} \\

\hline
\textbf{SN}& \textit{Single Stage Seeding} -- all seeds are activated in one stage at the beginning & $1$  \\ 
\hline
\textbf{SQ} & \textit{Sequential Seeding} -- one seed activation starts the stage and revives the previously stopped diffusion; each stage ends when diffusion stops & $n$  \\ 
\hline
\end{tabular}
\caption {Seeding strategies with $n$  number of seeds}
\label{tab:sq} 
\end{center}
\end {table*}

\begin{figure*}[!htb]
\centering
\centering
\includegraphics[width=\linewidth]{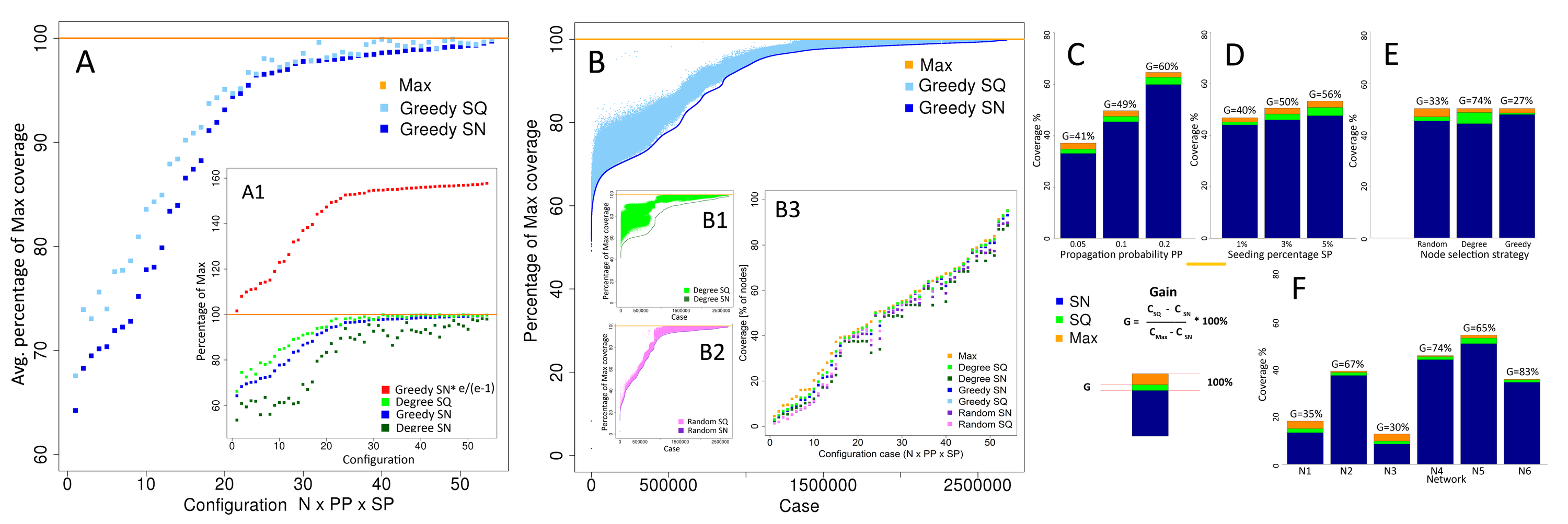}
\caption{Undirected networks: \textbf{(A)} The averaged performance of sequential SQ and single stage seeding SN with greedy based nodes selection as a fraction of the maximum coverage $C_{Max}$ and as a function of the network size N, probability of propagation PP across each edge, and the fraction of nodes selected as seeds (seed selection percentage) SP; see \textbf{(A1)} Performance of sequential SQ and single stage seeding SN with degree-based ranking in comparison with maximum coverage and the upper bound as a function of the individual configurations, each defined by N, PP, SP, and the random binary choice made at simulation initialization for each edge to propagate or not information across this edge; \textbf{(B)} Coverage of sequential method SQ as percentage of $C_{Max}$ placed between single stage seeding SN and Max for greedy nodes selection; \textbf{(B1)}, \textbf{(B2)} Sequential seeding performance SQ between single stage seeding SN and Max for random and degree-based node selection, respectively; \textbf{(B3)} Performance of sequential SQ and single stage seeding SN represented by percentage of activated nodes within network (coverage) for random seed selection, degree-based ranking and greedy seed selection in comparison with maximum coverage as a function of the individual configurations; \textbf{(C)} Gain for different propagation probabilities PP averaged over all cases; \textbf{(D)} Gain for different seeding percentages SP; \textbf{(E)} Gain for node ranking strategies based on random, greedy and degree selection; \textbf{(F)} Gain for networks N1 - N6. 
}

\label{fig:fig_2}
\end{figure*}

In the first phase, we ran simulations on six real complex networks with sizes in the range of 1.5k-17k nodes. The visualization of the networks and their basic properties are provided in the Supplementary Material. The independent cascades model (IC) \cite{Kempe:2003} was used with a given propagation probability $PP(a,b)$ that node $a$ activates (influences or infects) node $b$ in step $t+1$ under the condition that node $a$ was activated at time $t$ \cite{Wang:001}. The results for undirected networks assume that $PP(a,b)=PP(b,a)$ for all edges. This simplification allowed us to obtain the optimal seed set through the complete search even for largest networks among those used for experimentation. To explore the directed networks without this assumption, another large set of extensive experiments were performed on the smaller network, for which it was also possible to compute optimal solutions, see the next section.  

With the independent cascades model (IC) even a single seed can induce diffusion (which is crucial for sequential approaches), while in the linear threshold model \cite{Kempe:2003} (LT), a small seed set could have no effect. For that reason, experiments were carried out on the IC model.


Results achieved by sequential seeding (SQ) were compared with the single stage approach (SN) and the maximum coverage (Max) for the same setup: network and parameters: the propagation probability, the seeding percentage, and the seed ranking strategies: degree, greedy and random. The sequential coverage $C_{SQ}$ was on average 7.1\% better than the single stage coverage $C_{SN}$. 
The positive gain of SQ over SN was also confirmed by statistical tests, see Section Material and Methods. 

Results for greedy based ranking for 50,000 simulations represented as percentage of maximum coverage $C_{Max}$ averaged over individual configurations and ordered by the ratio $C_{SN}$ / $C_{Max}$ are presented in Fig.~\ref{fig:fig_2} (A). Additionally, the upper bound $C_{GreedySN} * e/(e-1)$ \cite{Kempe:2003} is for many configurations up to 50\% greater than real maximum value observed, see Fig.~\ref{fig:fig_2} (A1). These results demonstrate that upper limit derived from greedy approach is not tight. Sequential methods always outperform single stage ones, if we consider averaged coverage, for any strategy as well as for every configuration. Moreover, degree SQ is able to cover more nodes than greedy SN, especially for configurations with coverage significantly lower than $C_{Max}$, as shown in Fig.~\ref{fig:fig_2} (A1).

There were 8,100,000 individual simulation cases defined during simulation initialization, each case by its N, PP, SP, and binary random choice selected for each edge to propagate information or not across this edge. These cases are ordered by coverage obtained in the single stage method. The resulting plots demonstrate that $C_{SQ}$ performs better than $C_{SN}$ in almost every case. The greatest increase is observed for $C_{SN} \in [30\%,85\%]$; simply the space for improvement is larger in such cases.
 
The global results for all networks, strategies, parameters and random binary choices for each edge to propagate information or not across this edge made at the simulation initialization yield better values of $C_{SQ}$ than of $C_{SN}$ in 96.7\% of cases. The increase over 5\% was achieved in 20.2\% of cases. The results were dependent on node ranking strategy with 96.0\% better results for random rankings, 100\% for degree based rankings and 93.9\% for greedy approach based ranking. The improvement over 5\% was observed for 11.2\% cases for random strategy and as much as 38.7\% cases for degree-based selection and 10.9\% for greedy method. It should be also noted that, surprisingly, the random selection performs slightly better than the expensive greedy selection.

Results for individual cases with respect to maximum coverage are presented in Fig.~\ref{fig:fig_2} (B). The sequential  results are localized above the single stage border, filling the space towards maximum (100\%). The dispersion is the highest for degree based rankings Fig.~\ref{fig:fig_2} (B1). The lowest dispersion of results is observed for random based node selection, see Fig.~\ref{fig:fig_2} (B2). Moreover, if the single stage coverage is at least at the level of 90\%, all strategies are able to provide sequential cases very close to maximum. For lower coverage, only degree-based rankings can improve results so much.

The results revealed an important phenomenon: sequential seeding based on degree selection in 92.2\% of cases outperforms single stage greedy approach. A similar improvement is also observed for the averaged values. It means that sequential approach almost always is able to boost performance of the simple degree-based ranking over computationally expensive greedy heuristic. Moreover, $C_{degreeSQ}$ is greater than $C_{greedySQ}$ in 62.6\% of cases but for single stage methods such superiority can be observed in only 0.01\% of cases.

Hence, the main finding is that the computationally ineffective greedy strategy is suitable for single stage approach, while degree based selection for sequential seeding is significantly better than other selection methods. 

In general, the obtained results were dependent on the network profile and parameters of the diffusion. Space between the maximum coverage $C_{Max}$ and the single stage seeding $C_{SN}$ is an area in which sequential approaches deliver improvement. This area on average is $(C_{Max} - C_{SN})/C_{SN}$ so only 25\%, of the all simulation cases.

To evaluate the improvement, a gain measure $G$ was defined; it is based on average coverage values:  $G = ((C_{SQ} - C_{SN}) / (C_{Max} - C_{SN})) * 100\%$. It shows what part of the improvement area is reachable by sequential approach. Depending on process parameters and network, the gain varies from 30\% to 83\%. In general, the greater propagation probability value, the greater coverage and gain, see Fig.~\ref{fig:fig_2} (C). The same phenomenon arises for seeding percentages, as seen in Fig.~\ref{fig:fig_2} (D). Regarding node selection strategies, the highest average gain 74\% was observed for degree based selection, while for greedy and random strategies it is much lower: 27\% and 33\% respectively, as shown in Fig.~\ref{fig:fig_2}(E). The gain strongly depends on network structure, see Fig.~\ref{fig:fig_2} (F). The highest gain (83\%) was achieved for network N6, whereas the least gain (30\%) was achieved for network N3.

\section*{Results}
\section*{Experimental results for directed networks}

\begin{figure*}[!htb]
\centering
\begin{minipage}{1\textwidth}
\centering
\includegraphics[width=0.8\textwidth]{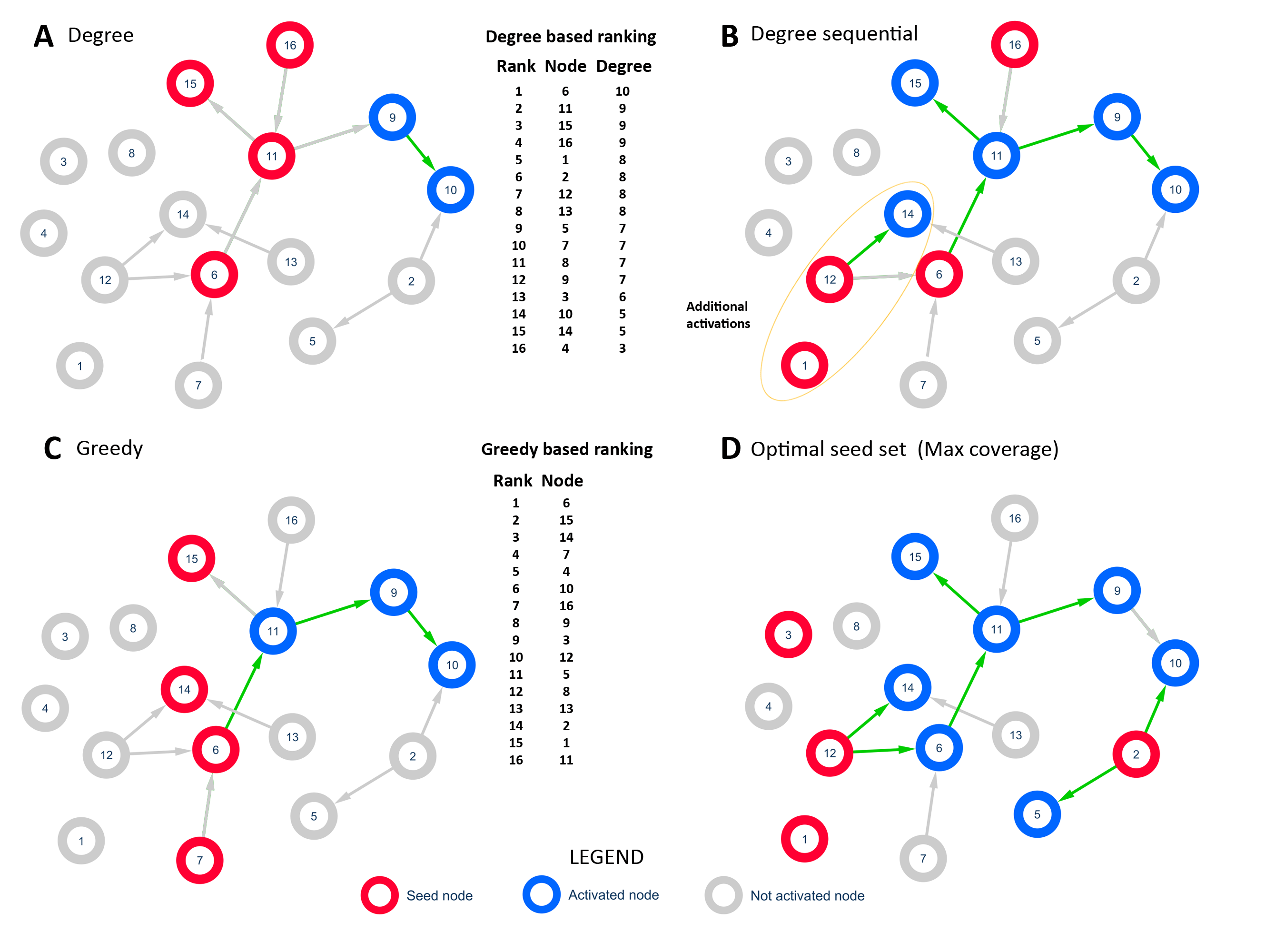}

\caption{One of the experimental cases for the directed network (run 4336) with the number of seeds $n=4$ and the propagation probability $PP=0.05$. For clarity and high network density only edges assigned the random choice of propagating information for them by coordinated execution are shown. Degree- and greedy-based rankings are computed for the full initial network with all edges. \textbf{(A)} Single stage seeding (SN) with degree-based ranking, the coverage $C_{SN}=6$, a diffusion cascade is visible when node 11 activates nodes 9 and 10. \textbf{(B)} Diffusion with sequential seeding and degree based selection; the ranking is the same as in (A); one seed is activated in each of four stages; in the first stage node 6 is activated according to its highest degree. It activates node 11 which in turn activates nodes 15 and 9; finally, node 9 activates node 10 and diffusion stops. The two seeds used in (A) are already activated by diffusion in this case. In the second stage node 16 is selected since it has the highest degree among not activated yet nodes. Its only neighbor, node 11, is already active so the process stops. In the next stage node 1 is selected as a seed but it lacks active edges so diffusion cannot progress. Fourth seed is node 12 and it activates node 14. Sequential seeding avoids using nodes 11 and 15 seeds which are activated as seeds in single stage seeding. This allow sequential seeding to activate two more seeds and three more nodes in total compared to single stage seeding. \textbf{(C)} Single stage seeding with greedy-based ranking and coverage $C_{SN}=7$. \textbf{(D)} Single stage seeding based on the seed set optimal for this individual case, including knowledge of which edges are active for propagation, with the resulting coverage $C_{SN}=11$. }
\label{fig:example}
\end{minipage}\hfill
\end{figure*}

\begin{figure*}[!htb]
\centering
\begin{minipage}{1\textwidth}
\centering
\includegraphics[width=\linewidth]{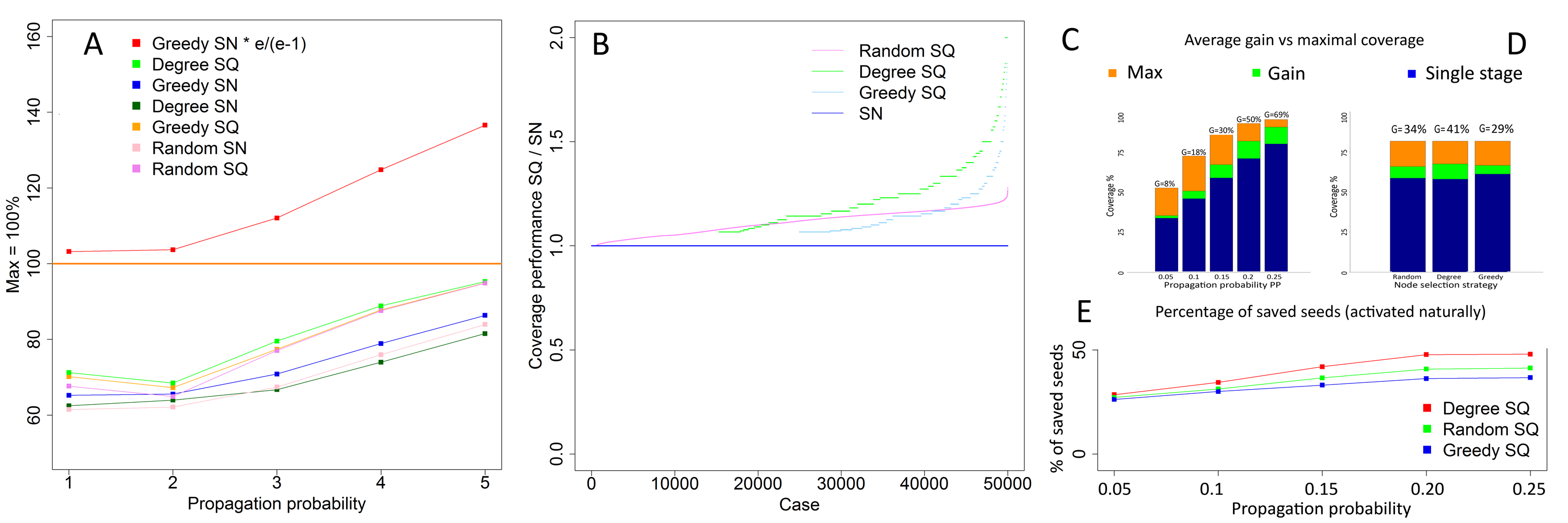}

\caption{The directed network: \textbf{(A)} Performance of sequential SQ and single stage seeding SN with degree-based ranking in comparison with maximum coverage; \textbf{(B)} The coverage achieved by sequential strategies in comparison to coverage obtained by a single stage approach; \textbf{(C)} Averaged gain for all used propagation probabilities; \textbf{(D)} Averaged gain for node ranking strategies based on random, greedy and degree selections; \textbf{(E)} Percentage of saved seeds for degree, random and greedy based selections.}

\label{fig:fig4}
\end{minipage}\hfill
\end{figure*}

As in case of undirected networks, coordinated execution experiments for a real directed network were run to analyze gain from sequential seeding compared to greedy approach and maximal possible coverage. Since they requires computationally expensive search for the optimal seed sets, only a small network with 16 nodes and 58 edges was used \cite{read1954cultures}. Following the coordinated execution principles, 10k instances of the network were randomly generated to assign binary choices of propagation or not for each directed edge, i.e., independently for $a \rightarrow b$ and $b \rightarrow a$ activation. One of these instances is presented in Fig~\ref{fig:example}.
For each instance and propagation probability PP, an optimal 4-node seed set was computed to estimate the maximum coverage for this instance and PP. Results for all probabilities and ranking strategies (degree, greedy, random) show that the highest increase of coverage was observed for the sequential degree based selection, see Fig.~\ref{fig:fig4} (B), (D). Moreover, the sequential approach with the degree rankings delivered better results than greedy-based selection regardless if used in single stage or sequential mode, as shown in Fig.~\ref{fig:fig4} (A), (D). It means that the degree approach should be among the first choices considered when selecting the node ranking method. 

The average performance (gain in coverage) is strongly dependent on propagation probabilities (PP) and it increases with PP values, see Fig.~\ref{fig:fig4} (C), even though higher probabilities leave less space for improvements, since they also raise the single stage coverage. This effect is present for all selection strategies. The reason is that although the processes with high PP reaches more than 80\% of nodes for all strategies and the area for gain is much smaller, the savings from the sequential approaches are even greater than for smaller PP's, as shown in Fig.~\ref{fig:fig4} (E),  benefiting the final results.

In general, the sequential approach enables us to save seeds and allocate them to other network regions. This gain varies for different propagation probabilities and ranking strategies, as shown in Fig.~\ref{fig:fig4} (E), and it comes from activation of seed candidates by inter-stage diffusion. The degree ranking reveals its superiority over other ranking methods also in seed saving, which may, in fact, be the reason why the sequential degree method outperforms the single stage greedy, as seen in Fig.~\ref{fig:fig4} (A). More than 25\% of seeds can be saved even for smallest propagation probabilities and up to 48\% for PP=0.25 and degree-based selection. These results may guide future research on methods to find the minimal number of seeds used sequentially to achieve the same coverage as in the single stage seeding. 

Overall, the main findings for directed networks are similar to the ones for undirected graphs: (1) sequential approach almost always significantly increases coverage of spreading and reaches beyond the current limits (never makes it worse), (2) sequential degree-based selection is commonly better than the greedy selection and (3) it is often close to the maximum, (4) usually, the greater propagation probability, the greater gain, and (5) sequential approach enables us to replace some of the initial seeds with the additional ones thereby increasing coverage, especially for greater PP and degree selection.

\section*{Discussion}

The presented results lay firm foundation for the earlier studies related to sequential seeding and address currently unanswered research questions ~\cite{jankowski2017balancing, jankowski2017dynamic, jankowski2017seeds}. Averaging the coverage obtained by the typical agent based simulations delivers improvement in about 90\% of the cases ~\cite{jankowski2017balancing}, the formal proof relying on the coordinated execution and corroborated by the new empirical studies shows that sequential seeding delivers at least as large coverage as the widely used single stage seeding; under modest assumption this coverage is provably larger. The performance of sequential seeding is compared with both greedy seed selection (which may be treated as the current limit) and the optimal seed set providing maximal possible coverage for a given number of seeds and the given network configuration. Experiments were performed for directed and undirected networks by means of independent cascades model~\cite{Kempe:2003}. Sequential usage of seeds selected by the simplest degree-based heuristics delivers better coverage than the greedy algorithm used in the single stage mode. However, coverage achieved by the greedy selection can also be further improved, if seeds are used sequentially instead of being all activated at the beginning of execution, even though the resulting coverage still remains worse than the one that the sequential degree selection delivers. 

Overall, the results confirm that sequential allocation of seeds improves the coverage of information diffusion in both directed and undirected complex networks. The  measure of improvements provided by this approach is showed. It helps to substitute seeding nodes activated by diffusion by selecting new seeds based on node's ranking which may activate new regions, see Fig.~\ref{fig:example} (B). 

As a result, sequential seeding commonly provides better coverage than the single stage approach, no matter which initial node selection strategy is used. In the worst case, the results remain the same but the experimental studies show that it happens in fewer than 10\% of all cases. 

For single stage mode the greedy approach is commonly better than typical ranking selections based on structural measures, e.g. degree and it can be treated as the method defining the current limit. However, it requires extensive prior simulations, which in practice are hardly possible due to (1) high computational complexity, and (2) feasibility of simulations for real data -- very rarely we can run thousands of diffusion processes when the results need to be delivered in the real-time of the campaign.

Sequential seeding outperforms the greedy search method. Moreover, sequential seeding using structural network measures like degree, for seed selection yields coverage greater than obtained by greedy, without necessity of any prior simulations. In our experiments, it happened in over 92\% of cases. It appears that using degree-based rankings with sequential strategy is better than using the computationally expensive greedy approach.

We observed that the main factor that defines performance of sequential seeding versus single stage seeding is to what extent seed selection method avoids selecting nodes that will be activated through diffusion. The gain of sequential seeding arises when the selected seed to be activated in the current stage is already active. In such a case, this seed is removed from the seed list and replaced by a highest ranking node which is not yet activated. It should be noted that we compare the resulting coverage with the real maximum coverage, which very often is much lower than the theoretical upper bound for the average maximum coverage suggested in \cite{Kempe:2003,kempe2015}, see red line in Fig.~\ref{fig:fig_2} (A1) and \ref{fig:fig4} (A).

Presented study has several implications for practice. Instead of introducing the product to large number of customers at the outset of commercial campaign, better strategy is seeding a small fraction of nodes and giving the chance to natural diffusion driven by social influence mechanisms to spread the content. Marketing budgets can be optimized if additional seeds are utilized only if campaign fails and revival is needed. Moreover, the knowledge gain from the initial spreading may improve seed selection for revival. Increased coverage of spreading might be crucial for campaigns with  limited budgets, such as spreading security information, or disease warnings and awareness. During massive campaigns habituation phenomenon can arise among customers resistant to marketing messages. It can be avoided by limiting the intensity of marketing activity. While campaigns with higher intensity can be perceived negatively as unsolicited massive communication, sequential strategies may avoid making such negative impact on customers. Sequential seeding is also a low risk strategy with possible high gains because, as we proved, results will never be worse when compared to single stage seeding.

\section*{Methods}

\subsection*{Independent cascade model}
We consider an independent cascade that is a stochastic diffusion model of information spread in the network initiated by seeds~\cite{Kempe:2003}. A basic, commonly used, single stage seeding (SN) consists of only one activation stage in which the fixed number of $n$ seeds are activated, see Table~\ref{tab:sq}. They initiate diffusion which runs until no more nodes can be activated. In the independent cascade model each diffusion step consists of a single attempt by all nodes activated in the previous step, to activate their direct not yet active neighbors with a given propagation probability (PP). We measure diffusion time in the number of such steps, assuming that each lasts a unit of execution time. 

\subsection*{Sequential seeding}

\begin{figure}
\centering
\includegraphics{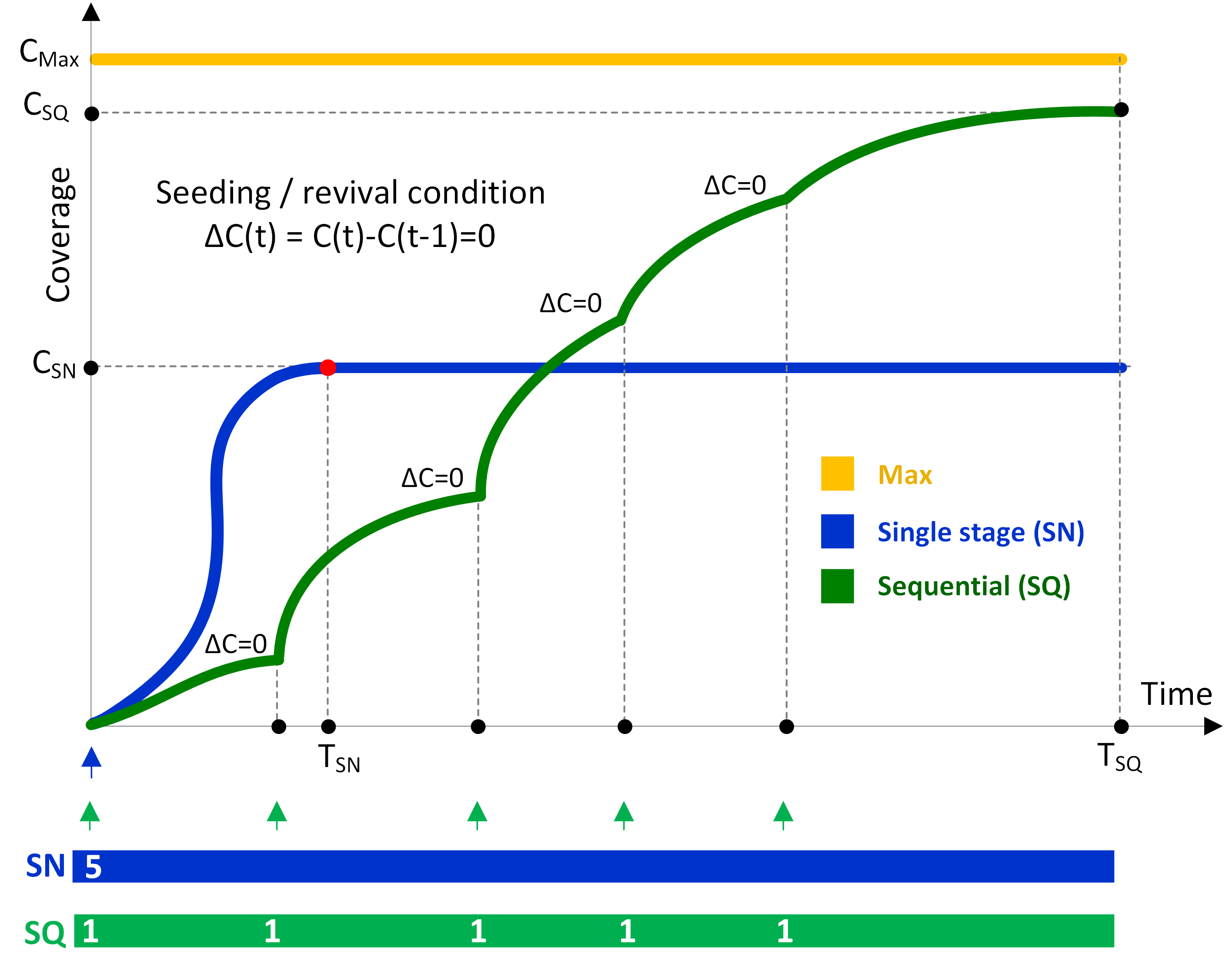}
\caption{Diffusion processes for the number of seeds $n=3$; the classical single stage method (SN) compared with sequential seeding strategy (SQ).}
\label{fig:SQ_SN_comp}
\end{figure}

Let $T_{SN}$ denote the time the whole process of the single stage seeding lasts, see Fig.~\ref{fig:SQ_SN_comp}. Often, some seeds activated at the beginning could have been naturally activated later on. The sequential seeding strategy (SQ) uses the same number of seeds $n$ as SN~\cite{jankowski2017balancing}. Its idea is to take the same initial ranking of nodes as in case of single stage seeding, but activate them sequentially in $n$ consecutive stages each with one seed activated. After each activation, the diffusion proceeds. The next activation is suspended until the current diffusion process stops, i.e., when the last diffusion step does not increase the coverage. In the independent cascade model, each recently activated node has only one chance to infect its neighbors. Hence, if the diffusion stops, the only way to continue activation is activating a seed. In the sequential method, it is the highest ranking node not yet activated. It means that nodes already activated by diffusion are omitted. The gain of sequential approach comes from the new areas activated by additional seeds substituting the already activated ones. Hence, the set of seeds activated in sequential seeding often differs from the one activated in the single stage approach. Sequential approach may be applied to any initial node ranking computed once at the beginning of the process. Such ranking may utilize random choice, any structural measures like commonly used degree or any other heuristic, e.g., greedy \cite{Kempe:2003, kempe2015}. Some questions arise here: what is the increased coverage $C_{SQ}$; is $C_{SQ}$ always greater than $C_{SN}$ and what is the gain achieved comparing with the estimate of the maximum coverage $C_{Max}$ for a given configuration. 

\subsection*{Coordinated execution and maximal coverage}
The need often arises to compare algorithms which during execution make random choices, e.g., to break ties, simulate random outcomes according to the given distribution, or are randomized in nature. In such cases often average values of performance are measured over many runs of each algorithm and used for comparison between them. Here, we propose coordinated execution as the more direct alternative, in which we compare performance of runs with the same choices made during executions with different algorithms, providing run by run comparison of the results. This requires that all random choices are made and recorded before each run for all compared algorithms. The subsequent runs for comparison of algorithms simply execute preprocessing of choices with unique seeds for random number generator.

When applied to comparing various seeding methods uses, the coordinated execution requires that each node activated in any execution makes the same edge diffusion transmission decision. To ensure that all edges randomly select their activity status before the coordinated executions are run for all variants of single stage and sequential strategies. It enables us to make fair comparisons between different seeding strategies.

Coordinated execution in the undirected network uses only one edge between two nodes. A propagation probability between two nodes $a \rightarrow b$ is the same as for $b \rightarrow a$ transmission. Coordinated execution in directed network replaces each edge with two directed edges, and then decides which directed edges will be activated in diffusion. The edge states are sampled from the binary set $\{0,1\}$ with state 1 being chosen with uniform probability PP. 

Edges assigned value of 0 are removed, while edges with value 1 stay, creating an instance of an active edge graph. Two instances of execution of sequential and single stage methods are coordinated if all edges originating from node activated in both instances have the same states. Such coordinated instances are used to compute the maximum coverage $C_{Max}$ for a given network and active edge choices. Such computation requires identification of all connected components in the active edge graph, which can be done in $O(e)$ steps, where $e$ is the number of active edges. Then, the maximum coverage $C_{Max}$ for a given seed size $n$ is equal to the total number of nodes within $n$ largest components.

Many instances of active edge graph are then created by simply using a different random number generator seed for each instance. The number of instances needed to collect the reliable estimate of the average coverage of the tested seeding methods is provided in \cite{kempe2015}. Computing the average of the maximum coverage over these instances provides a reliable estimate of an upper bound for the maximum average coverage of any seeding method. As shown in Fig.~\ref{fig:fig_2} (A1) and \ref{fig:fig4} (A), this upper bound is in most cases tighter than the one derived from the greedy seed selection. 

\subsection*{Formal proof of non-decreasing coverage}
In a formal proof, we show that the sequential seeding always provides at least as large coverage as the classical single stage approach does. Using coordinated execution, we also demonstrate that under modest assumptions the former approach is provably better than the latter when both use the same number of seeds and node ranking method. To do that, we demonstrate that there exist network configurations for which sequential seeding coverage is larger than the coverage yielded by the single stage seeding. Configurations are defined at simulation initialization by choosing  randomly the diffusion transmission status for each edge.

{\textbf Theorem:} For arbitrary node ranking and number of seeds, a sequential seeding execution has at least the same coverage $C_{SQ}$ as the corresponding single stage execution $C_{SN}$ coordinated with it. Moreover, if there is an initial seed $u$ which is reachable in the original graph from a seed $s$ with rank lower than rank of $u$, then there exists a configuration of the original graph for which the coverage for sequential seeding is larger than it is for single stage seeding.

{\textbf Proof:} For clarity, we assume that seeds are activated in the increasing order of their ranks. We start by observing that in sequential seeding, the seed ranked $r$ is activated no later than in stage $r$. It can be activated in the earlier stage either by diffusion or by having a seed with rank lower than $r$ activated by diffusion, thereby decreasing the ranks of all seeds ranked higher than the activated seed. If neither of these two cases happens before stage $r$ starts, in this stage, the seed ranked $r$ will be the highest ranking not yet activated seed and therefore it will be chosen for activation.

We observe also that not yet activated node will be activated by diffusion in a single stage execution of a given configuration if and only if this node is reachable from a seed by the edges active in this configuration. Indeed, if a node $m$ is reachable from a seed, this seed will be activated in the first and only stage of activation. Then, by the rules of diffusion, this seed's activation cascade will reach node $m$ and activate it, if it is still non-active. Conversely, the seed activated in step $t$ of diffusion in sequential seeding activates only not yet activated nodes that are distance $t$ from it. The upper bound for the number of diffusion steps before the diffusion stops is the diameter of the configuration. Hence, no node that is not reachable from any seed in the given configuration will be activated by diffusion. Since in sequential seeding all original seeds will be activated no later than by stage $n$, all nodes reachable from these seeds will also be activated by this stage, proving the first part of the Theorem.

Regarding the second part of the Theorem, let $H$ be a configuration in which all the edges on the reachability path from seed $s$ to node $u$ defined in the Theorem are active. At the end of stage $n-1$ of sequential execution of configuration $H$, all seeds activated in the single stage seeding are activated. Indeed by definition, all seeds ranked $n-1$ has to be activated by then, and the seed of rank $n$ has to be activated either by diffusion from seed $s$ when rank of $u$ is $n$ or by lowering of its rank by activation of seed $u$ from seed $s$ when rank of $u$ is less than $n$. Consequently, diffusion at stage $n-1$ comes from all seeds, and therefore all nodes activated in single stage seeding will be activated in sequential seeding by then. In the next stage $n$, not yet activated node will be activated as a seed, making the overall coverage of the sequential seeding larger for this configuration than the coverage of the single stage seeding; QED.

\subsection*{Greedy seed set search}
Greedy method requires many simulations to estimate the potential of a single node, i.e., its ability to activate other nodes via diffusion~\cite{Kempe:2003, kempe2015}. Here, the greedy method is based on approach presented in~\cite{kempe2015}. Averaged results of the greedy algorithm (coverage $C_{greedy}$ - the average total number of nodes activated by the process) are no worse than $C_{Max} * (1 - 1/e)$, where $C_{Max}$ is the expected maximum coverage~\cite{kempe2015}. It means that the greedy approach also defines the theoretical upper bound for maximum coverage: $C_{greedy} * e/(e-1)$. Due to its great coverage, the greedy algorithm is currently treated as the benchmark to beat and reference to. Since it requires prior simulations to compute the node potential, it is also very inefficient thus hardly applicable in practice, especially for large networks.

\subsection*{Experimental setup: undirected networks}
The experiments were carried out on six real complex networks: N1 - Condensed Matter collaboration \cite{newman2001scientific}, N2 - Communication network at University of California \cite{opsahl2009clustering}, N3 - High-Energy Theory collaboration network \cite{newman2001structure}, N4 - Political blogs \cite{adamic2005political}, N5 - ego-Facebook \cite{leskovec2012learning} and N6 - wiki-Vote \cite{leskovec2010predicting}. 

The parameters used in experimental configurations define diffusion, networks and seed selection strategy as shown in Table ~\ref{tab:par}. Three commonly used strategies were exploited: the highest degree, greedy \cite{kempe2015} and random selection. Simulation parameters create configuration space $N \times  PP \times  SP \times R$ with 162 configurations, each were independently applied to both single stage (SN) and sequential seeding strategies (SQ) using coordinated execution repeated 50,000 times, resulting in 162*50,000*2=16,200,000 simulation cases; 8,100,000 cases for each single stage and sequential strategies.

\begin {table}
\centering
\begin{tabular}{ l c l } 
\hline
\textbf{Parameter}& \textbf {Values} & \textbf {Variants} \\ 
\hline
Network - N & 6 & N1-N6 real networks\\ 
\hline
Propagation probability - PP& 3&0.05, 0.1, 0.20 \\
\hline
Seeds percentage - SP & 3&1\%, 3\%, 5\% \\ 
\hline
Seed ranking strategy - R & 3 & random, degree, greedy \\
\hline
\end{tabular}
\caption {Configuration space of the simulated diffusion processes}
\label{tab:par} 
\end {table}

For the greedy approach, finding the seed sets required another 10,000 simulations computed independently for each configuration, as defined in \cite{Kempe:2003, kempe2015}. The nodes were ranked according to their average coverage over all 10,000 simulations; a separate greedy ranking was computed for each of the 162 parameter combinations.

\subsection*{Statistical tests}
The positive gain of sequential approach (SQ) over single stage (SN) for experiments on undirected networks was also confirmed by the Wilcoxon signed rank test, with p\textless2.2e-16 and $\Delta$=1.9, with the Hodges-Lehmann estimator used as a difference measure. Values $\Delta>0$ demonstrate significantly larger coverage for $C_{SQ}$ than $C_{SN}$. Regarding seeding strategies based on node ranking methods, sequential approach increases coverage for random ranking on average by 3.2\%, with p\textless2.2e-16 and $\Delta$=1.6. Sequential seeding based on degree delivered coverage results 15.5\% better than single stage seeding, with p-value\textless2.2e-16 and $\Delta$=4.2. For sequential seeding and greedy based ranking 2.5\% average coverage improvement was achieved with p-value\textless2.2e-16 and $\Delta$=0.7.

\subsection*{Experimental setup: a directed network}

As much as 10,000 instances of coordinated execution for the small real network of 16 nodes were randomly selected, to ensure stability of the solutions; further instances would not affect results \cite{Kempe:2003}. Due to a small number of network nodes, only four seeds were used arbitrarily. Five propagation probabilities (PP) with values 0.05, 0.1, 0.15, 0.2 and 0.25 were applied. Similarly to the undirected version, three ranking strategies were analyzed: degree, greedy and random.


\onecolumn
\section*{Acknowledgment}
This work was partially supported by the National Science Centre, Poland, grant no. 2017/27/B/HS4/01216 (JJ), 2016/21/B/ST6/01463 (PK), 2015/17/D/ST6/04046 (RM), and 2016/21/D/ST6/02408 (PB), the European Union's Horizon 2020 research and innovation programme under the Marie Sk\l{}odowska-Curie grant agreement No. 691152 (RENOIR); the Polish Ministry of Science and Higher Education fund for supporting internationally co-financed projects in 2016–2019 (agreement no. 3628/H2020/2016/2); the Army Research Laboratory under Cooperative Agreement Number W911NF-09-2-0053 (the ARL Network Science CTA); and the Office of Naval Research Grant No. N00014-15-1-2640. 

\section*{Author contributions}
Author contributions: All authors contributed to the study design and manuscript preparation, BKS contributed proof of sequential seeding properties and the concept of coordinated execution, both JJ and RM conducted simulations and analyzed the data, PB prepared illustrative simulation cases. 

\section*{Competing interests}
Authors declare no conflict of interest.

\section*{Supplementary Information}
\label{sec:supplement}

\begin{figure*}[!htb]
\centering

\centering
\includegraphics[width=\linewidth]{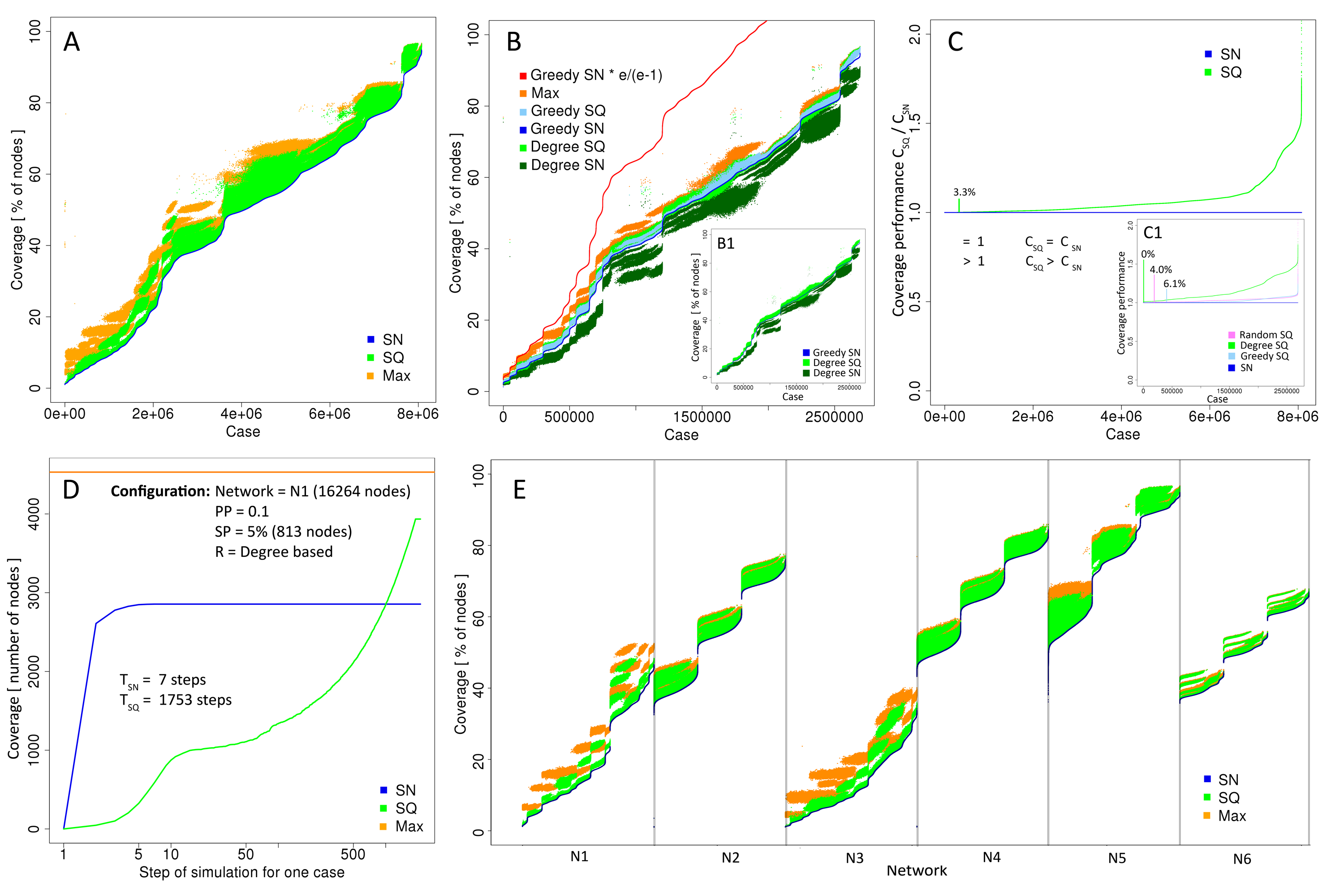}
\caption{\textbf{(A)} Coverage for sequential approach $C_{SQ}$ compared with single stage $C_{SN}$ and maximum coverage $C_{Max}$ for all configurations and all simulation cases; ordered by $C_{SN}$; \textbf{(B)} Coverage as percentage of all activated network nodes for greedy and degree-based single stage and sequential strategies and their relation to $C_{Max}$ and the upper bound; ordered by $C_{SN}$; \textbf{(B1)} Performance of sequential and single stage approaches with degree-based selection in relation to single stage greedy ranking;\textbf{(C)} Performance of sequential seeding SQ in the relation to single stage seeding SN, a vertical tick denotes percentage of cases with no gain: $C_{SQ}=C_{SN}$; \textbf{(C1)} Performance of random, degree and greedy based sequential seeding compared with single stage seeding SN; \textbf{(D)} Steps of one sequential and one single stage diffusion process in the coordinated execution for one configuration; \textbf{(E)} Coverage values for sequential $C_{SQ}$ and single stage seeding $C_{SN}$ as well as maximum coverage $C_{Max}$ for networks N1-N6 and all simulation cases.}
\label{fig:fig1supp}

\end{figure*}

\section*{Experimental results for undirected networks}

Each from 162 combinations from configuration space $N \times  PP \times  SP \times R$ was applied as a simulation configuration to single stage SN and sequential seeding SQ with the use of coordinated execution. Repeated simulations resulted in 16,200,000 simulation cases. Results from individual simulation cases are presented in Fig.~\ref{fig:fig1supp} (A). They show that $C_{SQ}$ performs better than $C_{SN}$ for most cases with coverage $C_{SQ}$ $\in [30\%,85\%]$.

Fig.~\ref{fig:fig1supp} (B) shows coverage for sequential approaches $C_{SQ}$ compared to single stage methods $C_{SN}$. The maximum coverage $C_{Max}$ for optimal seed set is presented as well as the upper bound $C_{GreedySN} * e/(e-1)$ for all cases. The results show that sequential seeding based on simple degree-based ranking in 92.2\% of cases outperforms single stage computationally expensive greedy heuristic, see Fig.~\ref{fig:fig1supp} (B1). Moreover, sequential seeding based on degree $C_{degreeSQ}$ is greater than sequential greedy approach $C_{greedySQ}$ in 62.6\% of cases.

For all simulation cases $C_{SQ}$ shows better results than $C_{SN}$ in 96.7\% of cases what as presented in Fig.~\ref{fig:fig1supp} (C). Increase above 5\% was achieved in 20.2\% of simulation cases. The obtained coverage was  dependent on seed selection strategy with 96.0\% better results for random selection, 100\% for degree based selection and 93.9\% for greedy based selection, see Fig.~\ref{fig:fig1supp} (C1). 

Fig.~\ref{fig:fig1supp} (D) shows one of real diffusion cases with visible differences between sequential (SQ) and single stage seeding (SN). As a result of better usage of natural diffusion processes sequential approach  outperforms the single stage approach.

Apart from user rankings the results were dependent on the network characteristics with visible differences seen in Fig.~\ref{fig:fig1supp} (E). Network N6 delivered the highest gain (83\%) while the least gain (30\%) was achieved for network N3.

\section*{Experimental results for directed network}

Performance of sequential seeding within directed network was analyzed for random node selection, degree based selection and greedy approach. Detailed results for all used probabilities and strategies for single stage and sequential seeding are presented in Table ~\ref{tab:table3}. The highest increase of coverage was observed for sequential degree based selection with results better than greedy selection for both single and sequential seeding. The highest gain for degree based selection was observed for low propagation probabilities 0.05, 0.01, 0.15. For higher probabilities differences between strategies are smaller.

\begin {table}
\centering
\begin{tabular}{ c c c c c c} 
\hline
\multicolumn{5}{c}{Coverage for random seed selection} \\
\hline
\textbf{PP}& \textbf {Single stage} & \textbf {\% of Max} & \textbf{Sequential} & \textbf{Increase} & \textbf {Gain} \\ 
\hline
0.05 & 5.43 & 62.2\% & 5.67 & 1.04 & 7.4\% \\ 
\hline
0.1	& 7.43 & 61.5\% & 8.17 & 1.09 & 16.0\% \\ 
\hline
0.15 & 9.65 & 67.4\% & 11.03 & 1.14 & 29.5\% \\ 
\hline
0.2	& 11.76 & 75.9\% & 13.56 & 1.15 & 48.3\% \\ 
\hline
0.25	& 13.34 & 83.9\% & 15.07 & 1.13 & 68.0\% \\ 
\hline

\multicolumn{5}{c}{Coverage for degree based seed selection} \\
\hline
\textbf{PP}& \textbf {Single stage} & \textbf {\% of Max} & \textbf{Sequential} & \textbf{Increase} & \textbf {Gain} \\ 
\hline

0.05 & 5.59 & 64.0\% & 5.99 & 1.07 & 12.5\% \\ 
0.1 & 7.54 & 62.4\% & 8.60 & 1.15 & 23.4\% \\ 
0.15 & 9.56 & 66.7\% & 11.39 & 1.21 & 38.5\% \\ 
0.2 & 11.45 & 73.9\% & 13.75 & 1.22 & 57.0\% \\ 
0.25 & 12.96&  81.5\% & 15.14 & 1.19 & 74.5\% \\

\multicolumn{5}{c}{Coverage for greedy seed selection} \\
\hline
\textbf{PP}& \textbf {Single stage} & \textbf {\% of Max} & \textbf{Sequential} & \textbf{Increase} & \textbf {Gain} \\ 
\hline

0.05 & 5.73 & 65.6\% & 5.88 & 1.02 & 5.0\% \\ 
0.1 & 7.88 & 65.2\% & 8.47 & 1.08 & 14.2\% \\ 
0.15 & 10.14 & 70.8\% & 11.08 & 1.09 & 22.4\% \\ 
0.2 & 12.22 & 78.9\% & 13.59 & 1.12 & 42.0\% \\ 
0.25 & 13.72 & 86.3\% & 15.09 & 1.11 & 62.9\% \\ 

\end{tabular}
\caption {Coverage for single stage and sequential seeding for used seed selection methods and propagation probabilities}
\label{tab:table3} 
\end {table}

\begin{table}[t]
\centering
\begin{tabular}{ c p{3.2cm} c c c c c c c} 
\hline
\textbf{No.} & \textbf{Network name} & \textbf{Ref} & \textbf {Nodes} & \textbf {Edges} & \textbf{Edge type} & \textbf{Components} & \textbf{CC} & \textbf {Diameter} \\ 
\hline
1 & N1 - Condensed Matter collaboration & \cite{newman2001scientific} & 16,264 & 47,594 & Undirected & 726 & 0.638 & 18 \\ \hline
2 & N2 - Communication network at University of California & \cite{opsahl2009clustering} & 1,899 & 20,296 & Undirected  & 4 & 0.109 & 8 \\ \hline
3 & N3 - High-Energy Theory collaboration network & \cite{newman2001structure} & 7,610 & 15,751 & Undirected & 581 & 0.486 & 19 \\ \hline
4 & N4 - Political blogs & \cite{adamic2005political} & 1,224 & 19,090 & Undirected & 2 & 0.320 & 8 \\ \hline
5 & N5 - ego-Facebook  & \cite{leskovec2012learning} & 4,039 & 88,234 & Undirected & 1 & 0.606 & 8 \\ \hline
6 & N6 - wiki-Vote & \cite{leskovec2010predicting} & 7,115 & 103,689 & Undirected & 24 & 0.141 & 7 \\ \hline \hline
7 & N7 - Social network of tribes & \cite{read1954cultures} & 16 & 114 & Directed & 1 & 0.519 & 3 \\ \hline

\end{tabular}
\caption {Description of  networks used in the experiments}
\label{tab:network stats} 
\end {table}

\begin{figure*}[!htb]
\centering
\includegraphics[width=\linewidth]{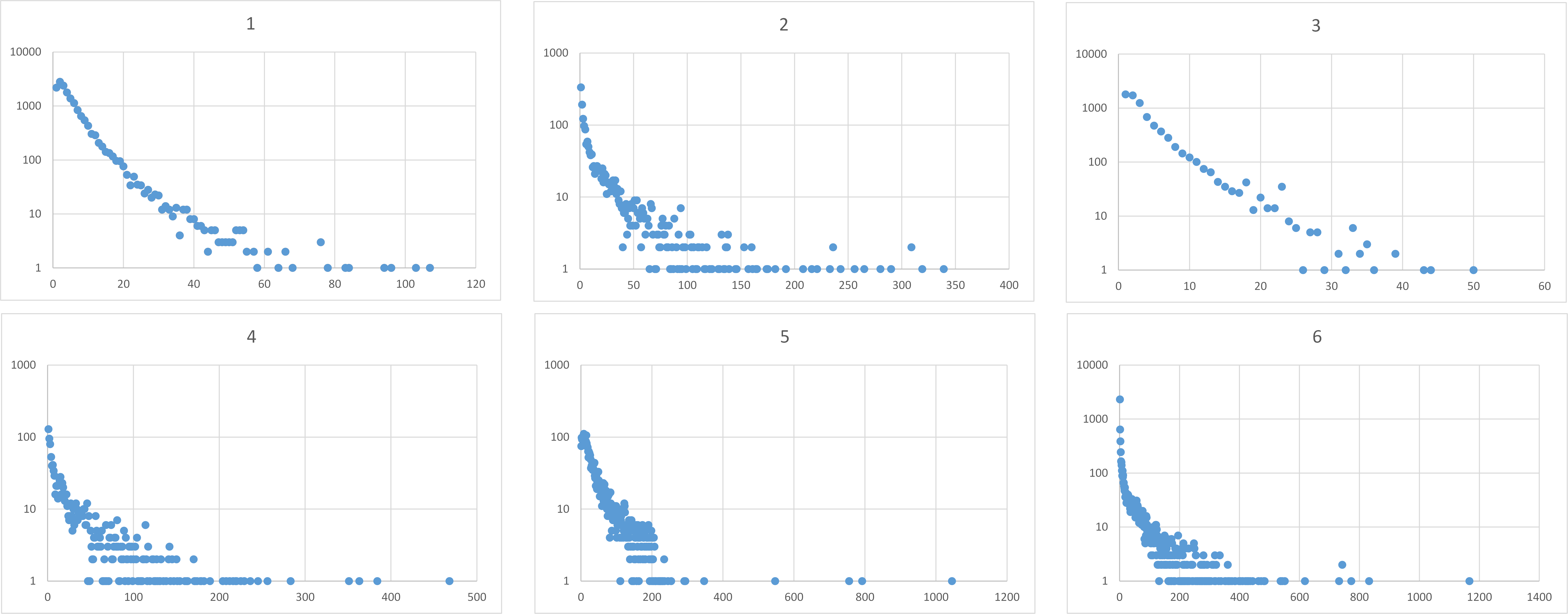}
\caption{Degree distribution for networks N1-N6}
\label{fig:degreeDist}

\end{figure*}

\end{document}